\documentclass[aps,preprint,amsmath,amssymb,showpacs,amsfonts]{revtex4}
\usepackage{epsfig}
\usepackage{graphicx}
\usepackage{dcolumn}
\usepackage{bm}
\usepackage{amsthm}
\usepackage{amsmath}

\newcommand{\beq}{\begin{equation}}
\newcommand{\eeq}{\end{equation}}
\newcommand{\bea}{\begin{eqnarray}}
\newcommand{\eea}{\end{eqnarray}}

\begin{document}
\begin{flushright}
AEI-2012-002
\end{flushright}
\bigskip
\bigskip

\title{The Relativistic Rindler Hydrodynamics}
\author{Christopher Eling$^1$}
\author{Adiel Meyer$^2$}
\author{Yaron Oz$^2$}

\affiliation{$^1$ Max Planck Institute for Gravitational Physics, Albert Einstein Institute, Potsdam 14476, Germany}
\affiliation{$^2$ School of Physics and Astronomy, Tel Aviv University, Tel Aviv 69978, Israel}

\date{\today}
\begin{abstract}
We consider a $(d+2)$-dimensional class of Lorentzian geometries holographically dual to a relativistic fluid flow in $(d+1)$ dimensions.
The fluid is defined on a $(d+1)$-dimensional time-like surface which is embedded in the $(d+2)$-dimensional bulk space-time and equipped with a flat intrinsic metric. We find two types of geometries that are solutions to the vacuum Einstein equations: the Rindler metric and the Taub plane symmetric vacuum. These correspond to dual perfect fluids with vanishing and negative energy densities respectively. While the Rindler geometry is characterized by a causal horizon, the Taub geometry has a timelike naked singularity, indicating pathological behavior. We construct the Rindler hydrodynamics up to second order in derivatives of the fluid variables and show the positivity of its entropy current divergence.

\end{abstract}


\maketitle

\tableofcontents

\newpage

\section{Introduction}

The holographic principle proposes that $(d+2)$-dimensional (quantum) gravitational theories are equivalent to $(d+1)$-dimensional field theories living on a boundary of the higher dimensional spacetime. The most concrete examples of holography are the so-called gauge/gravity correspondences, where quantum gravity with negative cosmological constant is dual to certain flat spacetime gauge theories. The gauge theory can be thought of as living on the timelike boundary at spatial infinity, in which the bulk spacetime is holographically encoded. A particularly interesting consequence of this duality is that the hydrodynamics of the gauge theory can be effectively described by the long time, long wavelength dynamics of a black hole living in the bulk. In this fluid-gravity correspondence, \cite{Bhattacharyya:2008jc,Bhattacharyya:2008kq} the Navier-Stokes equations of the fluid are equivalent to the subset of the General Relativity (GR) field equations called the momentum constraints, which constrain data on the timelike boundary surface.

However, by studying the dynamics of a Rindler acceleration horizon in flat spacetime \cite{Eling:2008af,nonrel,rel}, two of the authors argued that the relationship between holography and hydrodynamics is not limited to theories with negative cosmological constant. In the Rindler wedge of flat spacetime, the usual Minkowski vacuum is a thermal state at finite temperature (see for example \cite{Unruh:1983ac}). Just as in the AdS/CFT examples, one can study the long wavelength, long time perturbations of this background spacetime, which are expected to be dual to the hydrodynamics of the thermal state.  Indeed, recently \cite{Bredberg:2011jq,Compere:2011dx} it has been shown  that one can construct explicit bulk solutions to the vacuum Einstein equations dual to a particular non-relativistic fluid by perturbing around the Rindler geometry. The holographic fluid in this case is defined on an arbitrary timelike surface $S_c$ of fixed radial coordinate $r=r_c$ in the bulk geometry \cite{Bredberg:2010ky}. These are the hyperbolas associated with the worldlines of accelerated observers. Working in a non-relativistic hydrodynamic expansion, one can solve the field equations subject to the boundary conditions of a fixed flat induced metric on $S_c$ and a regular event horizon. The momentum constraints on $S_c$ again are the non-relativistic Navier-Stokes equations.

This result implies that there is an underlying duality between a field theory on $S_c$ and the bulk interior Rindler space. The nature of holography in asymptotically flat spacetimes has remained mysterious and it expected that the dual field theory will be nonlocal \cite{deHaro:2000wj}. An intriguing aspect of these results is that the dual fluid thermodynamics, constructed from the quasi-local Brown-York stress tensor, is characterized by zero equilibrium energy density even though there is non-zero temperature. Similarly, one can also show the entropy density of the fluid, equivalent to the area entropy density of horizon, is independent of its temperature.

Despite these unusual thermodynamical properties, the hydrodynamics of the Rindler fluid appears to be well-defined and contains no obvious pathologies. Studying this fluid in more detail may yield additional clues into the nature of the microscopic duality. In \cite{Compere:2011dx} the authors showed that their results can be obtained as the non-relativistic limit of an underlying relativistic fluid. In particular, they constructed a general theory for a viscous relativistic fluid with zero energy density and by matching to the non-relativistic solution were able to determine some of the viscous transport coefficients. While shear viscosity to entropy density ratio saturates the Kovtun-Son-Starinets bound of $1/4\pi$ \cite{Kovtun:2004de}, the bulk viscosity is not an independent transport coefficient even though the fluid is non-conformal. In the second order viscous hydrodynamics, there are six independent transport coefficients, but by matching to the non-relativistic solution one is only able to determine four of these. Later, \cite{Chirco:2011ex} studied how higher derivative corrections to the gravitational field equations affect the properties of the dual fluid. In the AdS/CFT correspondence, such terms are associated with quantum corrections or other deformations, which modify the values of the transport coefficients \cite{GBviscosity}. Interestingly, in this case only the second (and higher) order hydrodynamics is affected; the shear viscosity to entropy density ratio and the first order Navier-Stokes equations are universal.

In this paper, our main goal is to expand upon the results of \cite{Compere:2011dx} by completely determining the relativistic fluid dual to the Rindler spacetime. We start by considering a particular class of $(d+2)$-dimensional Lorentzian geometries. These metrics are stationary and on the slices $S_c$ the intrinsic metric is flat and the extrinsic curvature is such that the Brown-York quasi-local stress tensor has a perfect fluid form. Thus, these metrics can, in principle, act as the bulk gravitational dual to a fluid on $S_c$. Solving the vacuum Einstein equations, we find there are two possible branches of solutions. One is the Rindler solution described above and the other is the known as Taub plane symmetric vacuum and is associated with a fluid of negative energy density. The Taub metric has non-trivial curvature and, crucially, a naked singularity, which indicates pathological behavior in the dual field theory. In contrast, the Rindler solution is well-behaved, and using the formalism developed in \cite{Compere:2011dx} we find the bulk solution and fluid stress tensor in a derivative expansion up to second order, fixing the remaining two transport coefficients. In the fluid-gravity correspondence, the fluid entropy current is mapped into the area current \cite{Bhattacharyya:2008xc} of the evolving horizon surface. We compute this current to second order and find that its divergence is positive definite, consistent with Hawking's area theorem.

The plan of this paper is as follows. In Section II, we review the general construction of the solutions developed in \cite{Bredberg:2010ky,Bredberg:2011jq,Compere:2011dx} and describe the Rindler and Taub geometries. In Section III,  we present the earlier non-relativistic description of the Rindler fluid and use this to develop and eventually calculate the fully relativistic metric and its corresponding stress tensor. Section IV is devoted to the calculation of the holographic area entropy current and its divergence.
We then conclude with a brief discussion and description of possible future work.

\section{Ricci flat geometries and fluids}

In the following we will construct certain $(d+2)$-dimensional Lorentzian geometries holographically dual to a fluid flow in $(d+1)$ dimensions.
The fluid is defined on a $(d+1)$-dimensional timelike surface $S_c$ embedded in the $(d+2)$-dimensional bulk space-time. We choose the timelike surface to be defined by fixed bulk radial coordinate, $r=r_c$. Consider the following metric ansatz for the bulk geometry \cite{Bredberg:2010ky}
\begin{align}
ds^2 = g_{AB} dx^A dx^B = -h(r) dt^2 + 2 dt dr + e^{2\tau(r)} dx^i dx_i \ , \label{ansatz}
\end{align}
where $x^A = (t, x^i, r)$, $i=1..d$ and $d \geq 2$.
On surfaces of $r=r_c$, where $r_c$ is a constant, the induced metric is
\begin{align}
ds^2 =  \gamma_{\mu \nu} dx^\mu dx^\nu =  -h(r_c) dt^2 + e^{2\tau(r_c)} dx_i dx^i \ .
\end{align}
This metric is flat, as can be seen by the coordinate re-scaling $\bar{t} = \sqrt{h(r_c)}$ and $\bar{x}^i = e^{\tau(r_c)} x^i$, which leads to the standard Minkowskian form
\begin{align}
ds^2 = \gamma_{\bar{\mu} \bar{\nu}} d\bar{x}^\mu d\bar{x}^\nu= -d\bar{t}^2 + d\bar{x}_i d\bar{x}^i \ ,
\end{align}
where $\bar{x}^\mu = (\bar{t},\bar{x}^i)$

The Brown-York stress-energy tensor \cite{Brown:1992br} (in units where $16\pi G = 1$) associated with the $r=r_c$ slice is
\begin{align}
T_{\mu \nu} = 2 (K \gamma_{\mu \nu} - K_{\mu \nu}) \ , \label{BrownYork}
\end{align}
where $K_{\mu \nu} = \frac{1}{2} \mathcal{L}_N \gamma_{\mu \nu}$ and $\mathcal{L}_N$ is the Lie derivative along the normal to the slice $N^A$. Using (\ref{ansatz}) we find that
\begin{align}
T_{\bar{t} \bar{t}} = \rho = -2d \sqrt{h} \tau', \quad T_{\bar{i} \bar{j}} = p = 2\sqrt{h}\left((d-1) \tau'+\frac{h'}{h}\right) \delta_{\bar{i} \bar{j}} \ , \label{energypressure}
\end{align}
where primes represent derivatives with respect to $r$ and the expressions are evaluated at $r=r_c$. The stress-energy tensor has the form of a perfect fluid with energy density $\rho$ and pressure $p$.

We wish to determine whether there is a solution to the vacuum Einstein equations
\begin{align}
R_{AB} = 0 \ ,
\end{align}
of this general form. The Hamiltonian constraint, $R_{AB} N^A N^B$, where $N^A$ is the unit spacelike normal to the $r=r_c$ slices, is
\begin{align}
G_{AB} N^A N^B = \mathcal{R} - K_{\mu \nu} K^{\mu \nu} + K^2 \ ,
\end{align}
where $\mathcal{R}$ is the Ricci scalar associated with the induced metric of the slice. Re-expressing this equation in terms of the Brown-York stress tensor and using the fact that $\mathcal{R}=0$ for $r=r_c$, we get
\begin{align}
d T_{\mu \nu} T^{\mu \nu} = T^2 \ .
\end{align}
Inserting the general form of a perfect fluid stress tensor and (\ref{energypressure}), one finds that this condition is satisfied by two types of equations of state \cite{Compere:2011dx}
\begin{align}
(i)~~ \rho = 0,~~~~~~~~~~~(ii)~~ \rho = \frac{-2d}{(d-1)} p \ . \label{eos}
\end{align}

\subsection{The $\rho=0$ case: Rindler Geometry}

Consider first the equation of state $\rho=0$. Using Eqn. (\ref{energypressure}), this condition implies that $\tau'=0$ and as a result, $\tau= const.$. The remaining field equations imply that $h(r)=r$, and we get the metric
\beq
ds^2 = -r dt^2 + 2 dt dr + dx_i dx^i \label{Rindler} \ ,
\eeq
which describes a region of flat $(d+2)$-dimensional Minkowski space-time in ``ingoing Rindler" coordinates. The null surface $r=0$ acts as a causal horizon to accelerated observers, whose world-lines correspond to the surfaces of constant $r=r_c$.  Although the Rindler metric is just a patch of flat space-time, the associated quantum field theory on this background has many of the same properties as a black hole solution due to the existence of the causal horizon. In particular,  surfaces of $r=r_c$ have a local Unruh temperature (in units where $\hbar=c=1$)
\beq T = \frac{1}{4\pi \sqrt{r_c}} \ . \eeq
Strictly speaking, a Rindler horizon does not have a Bekenstein-Hawking entropy density. However, one can assign the Rindler horizon this entropy based on the holographic principle, or, more concretely, take the entropy to be the thermal entanglement entropy of the quantum fields in Rindler wedge \cite{entanglement}. This statistical entropy scales like an area, but is a UV divergent quantity. If a Planck scale cutoff is chosen appropriately, the entanglement entropy agrees with the Bekenstein-Hawking formula, i.e. in units where $16\pi G=1$
\beq s= 4\pi \ . \eeq
Given the existence of an equilibrium Unruh temperature and a Bekenstein-Hawking entropy density, the metric (\ref{Rindler}) may be considered
as providing a dual geometrical description of a perfect fluid with zero energy density in one lower space dimension. We will discuss the hydrodynamics of this case in detail in section III.

\subsection{The $\rho < 0$ case: Taub Geometry}

Consider next the second equation of state in (\ref{eos}). In order to construct the background, we plug-in the values of the energy density and pressure in terms of the metric functions into the equation of state, which gives
\begin{align}
\tau' + \frac{1}{d-1} \frac{h'}{h} = 0 \ . \label{feq1}
\end{align}
Consider the equation $R_{rr} = 0$. It yields
\begin{align}
\tau'^2 + \tau'' = 0 \ ,
\end{align}
which is solved by ,
\begin{align}
\tau(r) = \ln(C_1 r + C_2) \ .
\end{align}
Inserting this into (\ref{feq1}), we find
\begin{align}
h(r) = \frac{C_3}{(C_1 r + C_2)^{d-1}} \ .
\end{align}

Therefore the dual gravitational solution is
\begin{align}
ds^2 = - \frac{C_3}{(C_1 r+C_2)^{d-1}}  dt^2 + 2 dt dr + (C_1 r+ C_2)^2 dx_i dx^i \ .
\end{align}
Redefining the radial coordinate $\bar{r}= C_1 r + C_2$ and re-scaling the time coordinate, this metric takes the form
\begin{align}
ds^2 = - \frac{\mathcal{A}}{\bar{r}^{d-1}}  dt^2 + 2 dt d\bar{r} + \bar{r}^2 dx_i dx^i \ ,
\end{align}
where $\mathcal{A}$ is a constant.

In four-dimensions ($d=2$) this metric was found by A. H. Taub in 1951 \cite{Taub}. It can be considered the vacuum solution exterior to an infinite plane-symmetric object with uniform mass density. The Kretschmann scalar for this solution reads
\begin{align}
R_{ABCD} R^{ABCD} \sim \frac{1}{\bar{r}^{2(d+1)}} \ ,
\end{align}
which implies that there is a curvature singularity at $\bar{r}=0$ and the solution is asymptotically flat at infinity $\bar{r}=\infty$. The curvature singularity at $\bar{r}=0$ is timelike and naked, consistent with the fact that the energy density computed from the Brown-York stress tensor
\begin{align}
\rho = \frac{-2d\sqrt{\mathcal{A}}}{\bar{r}^{(d+1)/2}}
\end{align}
is always negative. The global structure of this metric was analyzed in \cite{Bedran:1996su}. At infinity there are two flat null surfaces, while the timelike naked singularity is located in the interior.
%
%

While it seems clear that this branch is problematic, let us nevertheless make a few remarks.

\noindent (i) One can make a spatial boost and re-write the metric in terms of the energy density $\rho$. This yields
\begin{align}
ds^2 = -\frac{\rho^2 \bar{r}^2}{4d^2} u_\mu u_\nu dx^\mu dx^\nu + 2 u_\mu dx^\mu d\bar{r} + \bar{r}^2 P_{\mu \nu} dx^\mu dx^\nu \ .
\end{align}
One could then formally allow the variables $\rho(x^\mu)$ and $u^\mu(x^\mu)$ and solve the field equations order by order in a derivative expansion in $\partial_\mu \rho$ and $\partial_\mu u_\nu$ as is done in the fluid-gravity correspondence \cite{Bhattacharyya:2008jc}. However, unlike the Rindler solution, in this case there is no casual horizon in the background zeroth order solution. Therefore, interpreting this geometry as being dual to a finite temperature state of a field theory (and then perturbations of this state as hydrodynamics) is problematic. For instance, imposing the thermodynamic identity $\rho + P = sT$ yields
\begin{align}
sT = \left(\frac{d+1}{1-d}\right) p \ .
\end{align}
Since $p > 0$, this equation implies formally a state of negative temperature (or negative entropy). A related fact is that the squared speed of sound $v_s^2 = \frac{dP}{d\rho}$ is negative, which indicates the dual field theory is characterized by an instability. It would be interesting to understand the role of this type of exotic solution in asymptotically flat holography.
\noindent
(ii) Following \cite{Compere:2011dx} we can create a scalar field Lagrangian that mimics the equation of state for the Taub metric. We consider
\begin{align}
I = \int d^{d+1} \sqrt{-g} F(X,\phi) \ ,
\end{align}
where $X = -(1/2) g^{\mu \nu} \partial_\mu \phi \partial_\nu \phi$. The stress-tensor is given by
\begin{align}
T_{\mu \nu} = -2 \frac{\partial F}{\partial g^{\mu \nu}} + g_{\mu \nu} F \ .
\end{align}
This gives
\begin{align}
T_{\mu \nu} = \frac{\partial F}{\partial X} \partial_\mu \phi \partial_\nu \phi + g_{\mu \nu} F \ .
\end{align}
If we identify the four-velocity (of a potential flow)
\begin{align}
u_\mu = \frac{\partial_\mu \phi}{2 \sqrt{X}} \ ,
\end{align}
the stress-tensor has the form of a perfect fluid with pressure $F$ and
\begin{align}
\rho = 2X \frac{\partial F}{\partial X}-F \ .
\end{align}
Imposing the equation of state, we find the condition on $F(X)$, which leads to an action of the form
\begin{align}
I = \int d^{d+1 x} \sqrt{-g} X^{-\frac{1}{2} (\frac{d+1}{d-1})} \ .
\end{align}

\section{The Rindler/Fluid Correspondence}

\subsection{General Setup}

In order to study the hydrodynamics of the fluid living on the $r=r_c$ slices in Rindler geometry, we have to perturb this background. The metric (\ref{Rindler}) corresponds to a fluid in thermodynamical equilibrium in the rest frame of some observer. The first step is to make a set of coordinate transformations to obtain a new metric corresponding to the fluid flow in an arbitrary frame. More geometrically, these transformations keep the induced metric at $r_c$ flat, in addition to preserving a perfect fluid form of the Brown-York stress energy tensor associated to the slice, the timelike Killing vector and the homogeneity in the $x^i$ direction \cite{Compere:2011dx}. One transformation is a translation of the radial coordinate, plus a re-scaling of $t$
\begin{align}
r \rightarrow r-r_h, \quad  t \rightarrow (1-r_h/r_c)^{-1/2} t \ ,
\end{align}
which moves the horizon from $r=0$ to an $r=r_h<r_c$. The second is a boost in the $x^i$ directions. The resulting metric can be written in the following manifestly boost covariant form
\begin{align}
ds^2 = g_{AB} dx^A dx^B = -(1+p^2(r-r_c)) u_\mu u_\nu dx^\mu dx^\nu - 2 p u_\mu dx^\mu dr + P_{\mu \nu} dx^\mu dx^\nu \ . \label{relativ1}
\end{align}
In this line element we replaced $r_h$ with the relativistic pressure $p$ using the formula
\beq
p = \frac{1}{\sqrt{r_c - r_h}} \ .
\eeq
The coordinates $x^\mu = (t,x^i)$. The fluid (boost) velocity is defined as $u^\mu = \gamma (1, v^i)$, where $\gamma = (r_c^2-v^2)^{-1/2}$ and $P_{\mu \nu} = \gamma_{\mu \nu} + u_\mu u_\nu$.

One can now investigate the hydrodynamic system dual to the above metric. To do that, we need to consider the dynamics of the metric perturbations within a hydrodynamic limit. In the literature \cite{Bredberg:2011jq, Compere:2011dx}, the first method developed was to work in terms of the non-relativistic variables $v^i$ and $r_h$ and allow these to be functions of space and time: $v^i(t, x^i)$ and $r_h(t, x^i)$, while $r_c$ remains fixed.  The metric is no longer flat and no longer a solution of the vacuum Einstein equation. One then introduces a particular non-relativistic hydrodynamical expansion, first proposed in \cite{Fouxon:2008tb,Bhattacharyya:2008kq}. In terms of a small parameter $\epsilon$,
\begin{align}
v^i \sim \epsilon v^i (\epsilon x^i , \epsilon^2 t) \quad P \sim \epsilon^2 P(\epsilon x^i , \epsilon^2 t) \ , \label{scaling}
\end{align}
where the \textit{non-relativistic} pressure $P(t,x^i)$ is defined as a small perturbation of the horizon radius
\beq
r_h = 0 + 2 P + O(\epsilon^4) \ .
\eeq
Expanding the metric (\ref{relativ1}) out to $O(\epsilon^2)$ in this manner yields the non-relativistic metric originally found by \cite{Bredberg:2011jq}
\begin{align}
d s^2 &= -r dt^2+2dt dr+d x_i d x^i \nonumber\\
&\quad -2\left(1-\frac{r}{r_c}\right)v_i d x^i dt -\frac{2v_i}{r_c} d x^i d r \nonumber\\
&\quad +\left(1-\frac{r}{r_c}\right)\Big[(v^2+2P) dt^2+\frac{v_iv_j}{r_c} d x^i d x^j\Big]+\left(\frac{v^2}{r_c}+\frac{2P}{r_c}\right)d t dr \ .
\label{seedmetric}
\end{align}
In GR, the momentum constraint equations on the surface $S_c$ are equivalent to the divergence of the Brown-York stress tensor
\beq R_{\mu A} N^A = \partial^\nu T^{BY}_{\mu \nu} \ , \label{momentumconstraint} \eeq
and constrain the fluid variables. At second and third order in $\epsilon$, momentum constraint equations are
\beq R^{(2,3)}_{\mu A} N^A = r_c^{-1/2} R^{(2,3)}_{t\mu} + r_c^{1/2} R^{(2,3)}_{r \mu} = 0 \ . \eeq
At second order, this is equivalent to the incompressibility condition $\partial_i v^i =0$.  At third order one finds the Navier-Stokes equations with a particular kinematic viscosity
\beq
\partial_t v_i + v^j \partial_j v_i + \partial_i P - r_c \partial^2 v_i = 0 \ . \label{NS}
\eeq
Imposing the incompressiblity condition implies (\ref{seedmetric}) is a solution to the Einstein equations up to $O(\epsilon^3)$. In \cite{Compere:2011dx} the higher order corrections to (\ref{seedmetric}) and the corresponding corrections to the incompressibility and Navier-Stokes equations were determined up to $O(\epsilon^6)$.

Comp\`{e}re, et. al  also noted that there should be a fully relativistic description of the fluid dual to Rindler spacetime. In this case the fluid variables the fluid velocity and pressure, which are allowed to depend on $x^\mu$: $u^\mu(x^\mu)$ and $p(x^\mu)$. They constructed the general form of a viscous fluid stress tensor to second order in derivatives with zero equilibrium energy density. The resulting stress tensor is
\begin{align}
T^{\mathrm{Rel}}_{\mu \nu} &= \rho u_\mu u_\nu + p P_{\mu \nu} - 2 \eta \mathcal{K}_{\mu \nu} \nonumber \\ & \quad + c_1 \mathcal{K}_\mu^\lambda \mathcal{K}_{\lambda \nu} + c_2 \mathcal{K}_{(\mu}^\lambda \Omega_{|\lambda|\nu)} + c_3 \Omega_\mu^{\,\,\,\lambda}\Omega_{\lambda \nu} +
c_4 P_\mu^\lambda P_\nu^\sigma D_\lambda D_\sigma \ln p \nonumber\\
& \quad  + c_5 \mathcal{K}_{\mu \nu}\,D\ln p + c_6 D^\perp_\mu \ln p \,D^\perp_\nu \ln p \ , \label{generalstress}
\end{align}
where $D \equiv u^\mu \partial_\mu$, $D^\perp_\mu = P^\nu_\mu \partial_\nu$, $\mathcal{K}_{\mu \nu} = P^\lambda_\mu P^\sigma_\nu \partial_{(\lambda} u_{\sigma)}$, and $\Omega_{\mu \nu} = P_\mu^\lambda P_\nu^\sigma\partial_{[\lambda}u_{\sigma]}$. The coefficient $\eta$ at first order is the usual shear viscosity. Note the absence of a bulk viscosity term in (\ref{generalstress}) at the same order. This is due to the fact that at viscous order we can impose the ideal order equation $\partial_\mu u^\mu = 0$, which follows from $\rho=0$. On the other hand, there can be viscous corrections to the energy density $\rho$, which can be parameterized very generally as
\begin{align}
\rho = b_1 \mathcal{K}_{\mu \nu}\mathcal{K}^{\mu \nu}+ b_2 \Omega_{\mu \nu}\Omega^{\mu \nu}+ b_3 D\ln p \,D\ln p + b_4 D^2 \ln p+b_5 D^\perp_\mu \ln p \,D^{\perp \mu}\ln p  \ .
\end{align}
The $c_i$, $i=1..6$, and $b_j$, $j=1..5$, are the possible new transport coefficients. Note that beyond equilibrium, the pressure and fluid velocity are no longer uniquely defined. This ambiguity is usually fixed by a choice of ``frame". Usually one works with the so-called Landau frame \cite{Landau}, which is constructed so that the viscous fluid velocity is defined as the velocity of energy transport $T^{(n)}_{\mu \sigma} u^\sigma = 0$ and $\rho$ does not receive viscous corrections. However, as we will see explicitly later on, this choice obviously turns out to be inconsistent in the Rindler fluid case at second viscous order. Instead one requires
\beq  T^{(n)}_{\mu \sigma} P^\mu_\nu u^\sigma = 0 \ , \label{Landau2}\eeq
and the pressure receives no corrections.

If one makes a non-relativistic expansion of (\ref{generalstress}) in terms of $v^i$ and $P$ and matches to the Brown-York stress tensor at $O(\epsilon^4)$ and $O(\epsilon^5)$ computed using the non-relativistic solution found in \cite{Compere:2011dx}, the values of the following transport coefficients can be read off
\begin{align}
\eta= 1, b_1 = -2\sqrt{r_c}, b_2 = 0, c_1=-2\sqrt{r_c}, c_3 = -4\sqrt{r_c}, c_2=c_4=-4\sqrt{r_c} \ .
\end{align}
However, to fix the remaining second order transport coefficients one has to work to even higher orders in the non-relativistic $\epsilon$ expansion. Instead we will take a more direct approach working with a relativistic hydrodynamic expansion.

\subsection{The relativistic fluid metric and stress tensor}

In this section we consider perturbations to the metric (\ref{relativ1}), treating $u^\mu(x^\mu)$ and $p(x^\mu)$ but leaving $r_c$ fixed. This follows the standard approach used in the fluid-gravity correspondence \cite{Bhattacharyya:2008jc}. Now the metric is
\begin{align}
ds^2 = g^{(0)}_{AB} dx^A dx^B =  -\Phi(x^\mu) u_\mu(x^\mu) u_\nu(x^\mu) dx^\mu dx^\nu - 2 p(x^\mu) u_\mu dx^\mu dr + P_{\mu \nu}(x^\mu) dx^\mu dx^\nu, \label{zerothorder}
\end{align}
where we have defined $\Phi = 1+p^2(r-r_c)$ for convenience. This metric is no longer a solution to the vacuum Einstein equations, but one can work order by order in an expansion in the derivatives of $u^\mu$ and $p$ with respect to $\partial_\mu$. This is equivalent physically to an expansion in a small, dimensionless Knudsen number $\lambda= \frac{\ell_{mfp}}{L}$, where $\ell_{mfp}$ is the mean free path of the underlying system and $L$ the characteristic scale of the perturbations to this system. With this solution in hand, we can compute the Brown-York stress tensor order by order. This stress tensor should have the form of (\ref{generalstress}), which will allow us to read off the transport coefficients in a direct way.

As a first step to illustrate how this works, we compute the Brown-York stress tensor for the metric (\ref{zerothorder}) at $r=r_c$. This is a solution at zeroth order i.e. $R_{AB} = 0 + O(\lambda)$. The space-like unit normal to this surface is
\begin{align}
n^r = \sqrt{g^{rr}}; \quad n^\mu = \frac{g^{r\mu}}{\sqrt{g^{rr}}} \ .
\end{align}
Therefore, we need to find the inverse metric (this is also needed in the our calculations later on). Using the formula $g^{AC} g_{CB} = \delta^A_B$, we find
\begin{align}
g^{rr} = p^{-2} \Phi; \quad g^{r\mu} = p^{-1} u^\mu; \quad g^{\mu \nu} = P^{\mu \nu} \ .
\end{align}
Thus, we find $n^r= p^{-1}$ and $n^\mu = u^\mu$. Using
\begin{align}
K_{\mu \nu}= \frac{1}{2}\left(n^A \nabla_A \gamma_{\mu \nu} + \gamma_{\mu \lambda} \partial_\nu n^\lambda + \gamma_{\lambda \nu} \partial_\mu n^\lambda \right) \ ,
\end{align}
(\ref{BrownYork}) gives
\begin{align}
T_{\mu \nu} dx^\mu dx^\nu = p P_{\mu \nu} dx^\mu dx^\nu \ ,
\end{align}
which as expected is the ideal part of (\ref{generalstress}) with $\rho=0$.

The strategy for solving the field equations is as follows. One introduces a first order correction to the metric $\delta g^{(1)}$,
\beq
 g = g^{(0)} + \delta g^{(1)} \ .
\eeq
The corrected metric at first order induces a $\delta R^{(1)}_{AB}$ at the same order (necessarily involving only radial derivatives). We want to solve for the metric $\delta g^{(1)}$ so that
\beq \delta R^{(1)}_{AB} + \hat{R}^{(1)}_{AB} = 0  \ ,\eeq
where $\hat{R}^{(1)}_{AB}$ comes from the zeroth order metric. This method can be generalized to solve for the metrics at higher order in $\lambda$. If we have a solution to $(n-1)$ order $g^{(n-1)}$, then one introduces a correction $\delta g^{(n)}$ so that
\beq \delta R^{(n)}_{AB} + \hat{R}^{(n)}_{AB} = 0 \ .  \label{fieldeqn} \eeq

The first step is to compute the general form of $\delta{R}^{(n)}_{AB}$. The Christoffel symbols are
\begin{align}
\delta \Gamma^{A}{}_{BC} = \frac{1}{2} g^{AD}_{(0)} \left( \bar{\nabla}_B \delta g^{(n)}_{CD} + \bar{\nabla}_C \delta g^{(n)}_{BD} - \bar{\nabla}_D \delta g^{(n)}_{BC}\right) \label{deltaG} \ ,
\end{align}
where $\bar{\nabla}_A$ is covariant derivative with respect to the background Rindler metric. Hence, we need the Christoffel symbols for the Rindler metric. These have the form
\begin{align}
\bar{\Gamma}^{r}{}_{\mu \nu} = \frac{1}{2} \Phi u_\mu u_\nu; \quad \bar{\Gamma}^{r}{}_{\mu r} = \frac{1}{2} p u_\mu; \quad \bar{\Gamma}^{\mu}{}_{\nu \lambda} = \frac{1}{2} p u^\mu u_\nu u_\lambda \
\end{align}
with the rest being zero.

In our solution, we will choose the gauge so that at all orders
\begin{align}
g_{rr} = 0; \quad g_{r \mu} = -p u_\mu \ .
\end{align}
This implies that $\delta g^{(n)}_{rr} = 0$ and $\delta g^{(n)}_{r \mu} = 0$. Expanding out (\ref{deltaG}) gives the following results:

\begin{align}
\delta \Gamma^{r}{}_{rr} &= 0 \\
\delta \Gamma^{r}{}_{\mu r} &= \frac{1}{2} p^{-1} u^\lambda \partial_r (\delta g^{(n)}_{\mu \lambda})\\
\delta \Gamma^{r}{}_{\mu \nu} &= -\frac{1}{2} p^{-2} \Phi \partial_r (\delta g^{(n)}_{\mu \nu}) - \frac{1}{2} u_\mu u_\nu u^\lambda u^\sigma \delta g^{(n)}_{\lambda \sigma}\\
\delta \Gamma^{\mu}{}_{rr} &= 0\\
\delta \Gamma^{\mu}{}_{r \nu} &= \frac{1}{2} P^{\mu \lambda} \partial_r (\delta g^{(n)}_{\nu \lambda})\\
\delta \Gamma^{\mu}{}_{\nu \lambda} &= \frac{1}{2}\left(-p^{-1} u^\mu \partial_r (\delta g^{(n)}_{\nu \lambda}) - p P^{\mu \tau} u_\nu u_\lambda u^\sigma \delta g^{(n)}_{\sigma \tau}\right)
\end{align}
Now we use the formula
\begin{align}
\delta R^{(n)}_{AB} = -\bar{\nabla}_A \delta \Gamma^{C}{}_{CB} + \bar{\nabla}_C \delta \Gamma^{C}{}_{AB} \ .
\end{align}
The final result is
\begin{align}
\delta R^{(n)}_{rr} &=  - \frac{1}{2} \partial^2_r (P^{\lambda \sigma} \delta g^{(n)}_{\lambda \sigma})\\
\delta R^{(n)}_{r \mu} &= \frac{1}{4} p u_\mu \partial_r (P^{\lambda \sigma} \delta g^{(n)}_{\lambda \sigma}) + \frac{1}{2} p^{-1} \partial^2_r( u^\lambda \delta g^{(n)}_{\mu \lambda})\\
\delta R^{(n)}_{\mu \nu} &= -\frac{1}{2} \left( u_\mu \partial_r (u^\lambda \delta g^{(n)}_{\nu \lambda}) + u_\nu \partial_r (u^\lambda \delta g^{(n)}_{\mu \lambda}))\right) - \frac{1}{2} \partial_r (\delta g^{(n)}_{\mu \nu}) - \frac{1}{2} p^{-2} \Phi  \partial^2_r (\delta g^{(n)}_{\mu \nu}) \nonumber \\ & - \frac{1}{2} u_\mu u_\nu \partial_r (u^\lambda u^\sigma \delta g^{(n)}_{\lambda \sigma}) + \frac{1}{4} \Phi u_\mu u_\nu \partial_r (P^{\lambda \sigma} \delta g^{(n)}_{\lambda \sigma}).
\end{align}
Notice that these satisfy
\beq \delta R^{(n)}_{\mu A} n^A = 0 \ . \label{constraint0} \eeq
Using (\ref{fieldeqn}) we can now obtain the general solution for $\delta g^{(n)}_{AB}$. This solution needs to be consistent with the following boundary conditions: (i) no singularity appearing at $r=0$ and (ii) induced metric remains flat on $r=r_c$. The second implies that all the $n \geq 1$ order corrections must vanish at $r=r_c$. Projecting into components normal and transverse to $u^\mu$ we find
\begin{align}
P_\mu^\lambda P_\nu^\sigma \delta g^{(n)}_{\lambda \sigma} &= 2 p^2 \int^{r}_{r_c} \frac{1}{\Phi} dr' \int^{r'}_{r_c - \frac{1}{p^2}} P_\mu^\lambda P_\nu^\sigma \hat{R}^{(n)}_{\lambda \sigma} dr'' \label{gensol1} \ , \\
u^\lambda P^\sigma_\mu \delta g^{(n)}_{\lambda \sigma} &= (1/2) (1-r/r_c) V^{(n)}_\mu(x) - 2 p \int^{r_c}_{r} dr' \int^{r_c}_{r'} dr'' P^\lambda_\mu \hat{R}^{(n)}_{r \lambda} \label{gensol2} \ , \\
u^\lambda u^\sigma \delta g^{(n)}_{\lambda \sigma} &= (1-r/r_c) A^{(n)}(x) + p \int^{r_c}_{r} dr' \int^{r_c}_{r'} dr'' \left(p P^{\lambda \sigma} \hat{R}^{(n)}_{\lambda \sigma} - p^{-1} \Phi \hat{R}^{(n)}_{rr} - 2 \hat{R}^{(n)}_{r \lambda} u^\lambda \right). \label{gensol3}
\end{align}
where $V^{(n)}_\mu$ ($V^{(n)}_\mu u^\mu = 0$) and $A^{(n)}$ are free, undetermined functions at this stage.

\subsubsection{The first viscous order}

With this general solution for any $n$, we now consider the first order solution, which requires the $\hat{R}^{(1)}_{AB}$ computed from the zeroth order equilibrium Rindler metric. To start, we need the Christoffel symbols to first order in $\lambda$. These can be read off from the second order connection presented in Appendix A.
%

Using these results, we find ultimately for Ricci tensor components
\begin{align}
\label{hattedRicci}
\hat{R}_{rr}^{(1)} &= 0 \nonumber \\
\hat{R}_{r \mu}^{(1)} &= 0 \nonumber \\
\hat{R}_{\mu \nu}^{(1)} &= \partial_{(\mu} p u_{\nu)} + D p ~u_\mu u_\nu + \frac{1}{2} p (\partial_\lambda u^\lambda) u_\mu u_\nu + p u_{(\mu} a_{\nu)} \ ,
\end{align}
where the operator $D = u^\lambda \partial_\lambda$ and $a_\mu = u^\nu \partial_\nu u_\mu$.

%
The equation $R_{\mu A} n^A = 0$ is the momentum constraint. However, as we saw earlier (\ref{constraint0}) the $\delta R_{\mu A}$ piece satisfies this condition automatically. Therefore we must have $u^\nu \hat{R}_{\mu \nu} =0$. Projecting along $u^\mu$ and orthogonal to $u^\mu$ with the projector, one finds the relativistic ideal hydro equations
\begin{align}
\partial_\mu u^\mu &= 0 \ , \\
a_{\mu} + P^\lambda_\mu \partial_\lambda \ln p &= 0 \ .
\end{align}
So, as expected, the momentum constraints are the relativistic Navier-Stokes equations.

%

Inserting (\ref{hattedRicci}) into (\ref{gensol1}-\ref{gensol3}) yields the solution
\begin{align}
u^\mu P^\lambda_\nu \delta g^{(1)}_{\mu \lambda} &= (1/2) (1-r/r_c) V^{(1)}_\nu(x^\mu) \ , \\
P^\lambda_\mu P^\sigma_\nu \delta g^{(1)}_{\lambda \sigma} &= 0 \label{eq:PPg} \ , \\
u^\mu u^\nu \delta g^{(1)}_{\mu \nu}  &= (1-r/r_c) A^{(1)}(x^\mu) \ . \label{eq:uug}
\end{align}
%
%
%
To determine these functions, we need to impose the ``frame" conditions on the 1st order, viscous part of the Brown-York stress tensor. Using (\ref{BrownYork}) we find the first order part of the stress tensor is
\begin{align}
T^{(1)}_{\mu \nu} &= \left((r_c p)^{-1} A^{(1)} + 2 p^{-1} Dp  \right) \gamma_{\mu \nu} + (r_c p)^{-1}(A^{(1)} u_\mu u_\nu - u_{(\mu} V^{(1)}_{\nu)})  \nonumber \\
& - 2 \partial_{(\mu} u_{\nu)} - 2 p^{-1} u_{(\mu} \partial_{\nu)} p,
\end{align}
where we have imposed $\partial_\mu u^\mu = 0$. We now require the stress tensor satisfies the Landau-like condition (\ref{Landau2}).
%
%
This yields an equation for $V^{(1)}_\mu$:
\begin{align}
V^{(1)}_\mu = - 2 r_c P^\nu_\mu \partial_\nu p + 2 r_c p a_\mu \ .
\end{align}
The second condition we demand is for $p$ to be the pressure at all viscous orders. Thus the term proportional to $\gamma_{\mu \nu}$ must vanish. This implies
\begin{align}
A^{(1)} = -2 r_c p D(\ln p) \ .
\end{align}
Feeding these results back into the stress tensor, we find simply
\beq T^{(1)}_{\mu \nu} = - 2 \mathcal{K}_{\mu \nu} \ . \eeq
So, as we expected, the shear viscosity $\eta = 1$. The complete solution for the metric to first order is (imposing the ideal hydrodynamics equations)
\begin{align}
ds^2 &= -(1+p^2(r-r_c)) u_\mu u_\nu dx^\mu dx^\nu - 2 p u_\mu dx^\mu dr + P_{\mu \nu} dx^\mu dx^\nu \nonumber \\
&+ 2 p (r-r_c) D(\ln p) u_\mu u_\nu  dx^\mu dx^\nu - 4 p (r-r_c)  u_{(\mu} P^{\lambda}_{\nu)} \partial_\lambda \ln p  dx^\mu dx^\nu  \ . \label{metric1}
\end{align}

\subsubsection{The second viscous order}

In order to solve to second order in $\lambda$, we need to find $\hat{R}^{(2)}_{AB}$. The first step is to compute the connections out to second order. Then it's a matter of grinding though the calculations of the Ricci tensor, determining the solution, and then computing the Brown-York stress tensor. This will fix all of the second order $c_i$ transport coefficients. The connections and the Ricci tensor that come from metric (\ref{metric1}) are complicated, so we present them in Appendix A.

We find that at second order the momentum constraint equations $R_{\mu A} n^A = 0$ projected once on $u^\mu$ and once on $P^\mu_\sigma$ are:
\begin{align}
\partial_\mu u^\mu - p^{-1} \partial_\sigma u^\lambda \partial_\lambda u^\sigma - p^{-1} P^{\rho \sigma} \partial_\rho u^\nu \partial_\sigma u_\nu&= 0 \\
a_{\mu} + P^\lambda_\mu \partial_\lambda \ln p - p^{-1} P^{\nu \sigma} \partial_\nu \partial_\sigma u_\mu + p^{-1} u_\mu P^{\nu \sigma} \partial_\nu u^\lambda \partial_\sigma u_\lambda +2 p^{-1} P^{\nu \sigma} \partial_\nu u_\mu \partial_\sigma \text{ln} p &= 0.
\end{align}

Solving the equations (\ref{gensol1}-\ref{gensol3}) for $\delta g_{\mu \nu} ^{(2)}$ we get:
\begin{align}
u^\mu P^\lambda_\nu \delta g^{(2)}_{\mu \lambda} &= \frac{1}{2} p^2 (r-r_c)^2 \left( -2 P^\mu_\nu \partial_\mu u^\rho \partial_\rho \text{ln} p  + 2 P^{\lambda \rho} \partial_\lambda u_\nu \partial_\rho \text{ln} p- P^{\rho \sigma} \partial_\rho \partial_\sigma u_\nu + u_\nu P^{\rho \sigma} \partial_\sigma u_\mu \partial_\rho u^\mu \right) \nonumber \\
&+ \frac{1}{2} (1-\frac{r}{r_c}) V^{(2)}_\nu(x^\mu) \\
P^\lambda_\mu P^\sigma_\nu \delta g^{(2)}_{\lambda \sigma} &= (r-r_c) P^\alpha_\mu P^\beta_\nu \left(- \partial_\alpha \text{ln} p \partial_\beta \text{ln} p+ 2 \partial_\alpha  \partial_\beta \text{ln} p + 2 \partial_{(\alpha} u_{\beta)} u^\rho \partial_\rho \text{ln} p - 2 u^\sigma \partial_\sigma \partial_{(\alpha} u_{\beta)} \right. \nonumber \\
&\hspace{75pt} \left. + 2 u^\sigma \partial_\sigma u_{(\beta} \partial_{\alpha)} \text{ln} p -a_\beta a_\alpha - \frac{3}{2} \partial_\alpha u^\lambda \partial_\beta u_\lambda + \frac{1}{2} P^{\sigma \rho} \partial_\sigma u_\beta \partial_\rho u_\alpha +\partial_\sigma u_{(\alpha} \partial_{\beta)} u^\sigma\right) \nonumber \\
&+ \frac{1}{4} p^2 (r-r_c)^2 P^\alpha_\mu P^\beta_\nu \left(- 2 \partial_\sigma u_{(\alpha} \partial_{\beta)} u^\sigma + P^{\sigma \lambda} \partial_\sigma u_\beta \partial_\lambda u_\alpha + \partial_\alpha u^\lambda \partial_\beta u_\lambda \right) \\
u^\mu u^\nu \delta g^{(2)}_{\mu \nu}  &=\frac{1}{4} p^2 (r-r_c)^2 \left(\partial_\sigma u^\alpha \partial_\alpha u^\sigma + P^{\sigma \lambda} \partial_\sigma u^\nu \partial_\lambda u_\nu + 4 P^{\sigma \lambda} \partial_\sigma \text{ln} p \partial_\lambda \text{ln} p \right) \nonumber \\
&+ \frac{1}{4} p^4 (r-r_c)^3 \left( -\partial_\sigma u^\alpha \partial_\alpha u^\sigma + P^{\sigma \lambda} \partial_\sigma u_\beta \partial_\lambda u^\beta \right) + (1-\frac{r}{r_c}) A^{(2)}(x^\mu).
\end{align}
The corresponding Brown-York stress energy tensor is:
\begin{align}
T_{\mu \nu} &=\gamma_{\mu \nu} \left(p + p^{-1} \left(r^{-1}_cA^{(2)} + \partial_\sigma u^\lambda \partial_\lambda u^\sigma + P^{\rho \sigma} \partial_\rho u^\nu \partial_\sigma u_\nu \right) \right) + p u_\mu u_\nu \nonumber \\
&+ \left( - 2 \partial_{(\mu)} u_{\nu)} + 2 u_{(\mu} \partial_{\nu)} \text{ln} p + 2 u_\mu u_\nu u^\lambda \partial_\lambda \text{ln} p \right) \nonumber \\
&+ r^{-1}_c p^{-1} u_\mu u_\nu A^{(2)} - r^{-1}_c p^{-1} u_{(\mu} P^\rho_{\nu)} V^{(2)}_\rho \nonumber \\
&- p^{-1} P^\alpha_\mu P^\beta_\nu \left(- 4 \partial_\alpha \text{ln} p \partial_\beta  \text{ln} p + 2 \partial_\alpha  \partial_\beta \text{ln} p + 2 \partial_{(\alpha} u_{\beta)} u^\rho \partial_\rho \text{ln} p - 2 u^\sigma \partial_\sigma \partial_{(\alpha} u_{\beta)} \right. \nonumber \\
&\hspace{60pt} \left.  - \frac{3}{2} \partial_\alpha u^\lambda \partial_\beta u_\lambda + \frac{1}{2} P^{\sigma \rho} \partial_\sigma u_\beta \partial_\rho u_\alpha +\partial_\sigma u_{(\alpha} \partial_{\beta)} u^\sigma\right). \label{BY2nd}
\end{align}
As before, we impose the frame condition that the pressure be unchanged from its equilibrium value. This condition eliminates the derivative parts of first term proportional to $\gamma_{\mu \nu}$ in (\ref{BY2nd}) and fixes $A^{(2)}$ to a non-zero value. Note however that the stress tensor now has a term proportional to $u_\mu u_\nu$ (in the third line above). Hence at this point we see the traditional Landau frame $T^{(2)}_{\mu \nu} u^\mu = 0$ will be inconsistent. To fix $V^{(2)}_\mu$ we instead require (\ref{Landau2}).

With these values fixed, the stress tensor can be put in a more conventional form:
\begin{align}
T_{\mu \nu} &=pP_{\mu \nu} -2 \mathcal{K}_{\mu\nu} - 2 p^{-1} u_\mu u_\nu \ \mathcal{K}_{\alpha \beta}  \mathcal{K}^{\alpha \beta} - 2p^{-1} \mathcal{K}_{\mu \rho} \mathcal{K}^\rho_\nu  - 4 p^{-1} \mathcal{K}^\rho_{(\mu}  \Omega_{|\rho|\nu)}  - 4 p^{-1} \Omega_{\mu \rho} \Omega^\rho_{\hspace{3pt} \nu}
\nonumber \\
&  - 4 p^{-1} P^\alpha_\mu P^\beta_\nu \partial_\alpha \partial_\beta \text{ln} p  - 4 p^{-1} \mathcal{K}_{\mu \nu} D \text{ln} p  + 4 p^{-1} D^\perp_\mu \text{ln} p D^\perp_\nu \text{ln} p
\end{align}
Due to the $u_\mu u_\nu$ term, one can see that the energy density is no longer zero. It gets following correction in the second order:
\begin{align}
\rho &= T_{\mu \nu} u^\mu u^\nu = - p^{-1} \partial_\sigma u^\lambda \partial_\lambda u^\sigma  - p^{-1} P^{\rho \sigma} \partial_\rho u^\nu \partial_\sigma u_\nu = -\frac{2}{p} \mathcal{K}_{\mu \nu}  \mathcal{K}^{\mu \nu}.
\end{align}
We can also read of from the stress tensor the transport coefficients. In the notation of \cite{Compere:2011dx}, we get:
\begin{align}
c_1 = -2 p^{-1}, \hspace{15pt} c_2 = c_3 = c_4 = c_5 = -4 p^{-1},  \hspace{15pt} c_6 = 4 p^{-1}.
\end{align}
which is in agreement with \cite{Compere:2011dx}, who found the first four transport coefficients and the energy density. \\
Finally for the metric solution $\delta g_{\mu \nu} ^{(2)} $ we get:
\begin{align} \label{second order metric}
\delta g^{(2)}_{\mu \nu} &= u_\mu u_\nu \left(\frac{1}{2} p^2 (r-r_c)^2 \left(\mathcal{K}_{\alpha\beta} \mathcal{K}^{\alpha\beta} + 2 P^{\sigma \lambda} \partial_\sigma \text{ln} p \partial_\lambda \text{ln} p \right) \right. \nonumber \\
& \hspace{40pt} + \left. \frac{1}{2} p^4 (r-r_c)^3 \left( \Omega_{\alpha\beta} \Omega^{\alpha\beta} \right) + 2 (r-r_c) \left(\mathcal{K}_{\alpha\beta} \mathcal{K}^{\alpha\beta} \right) \right) \nonumber \\
&+ 2 u_{(\mu} P^\rho_{\nu)} \left( \frac{1}{2} p^2 (r-r_c)^2 \left( 4 \Omega_\rho ^{\hspace{3pt} \sigma} \partial_\sigma \text{ln} p + P^{\lambda \sigma} \partial_\lambda \partial_\sigma u_\rho \right) \right. \nonumber \\
& \hspace{45pt} \left.- (r-r_c) ( - P^{\lambda \sigma} \partial_\lambda \partial_\sigma u_\rho +  2\mathcal{K}_\rho ^\sigma \partial_\sigma \text{ln} p - 2\Omega_\rho ^{\hspace{3pt} \sigma} \partial_\sigma \text{ln} p) \vphantom{\frac{1}{2}} \right) \nonumber \\
&+ (r-r_c) \left( 2 \mathcal{K}_{\mu \rho} \mathcal{K}^\rho_\nu  + 4 \mathcal{K}^\rho_{(\mu}  \Omega_{|\rho|\nu)}  + 4  \Omega_{\mu \rho} \Omega^\rho_{\hspace{3pt} \nu}
 + 4  P^\alpha_\mu P^\beta_\nu\partial_\alpha  \partial_\beta \text{ln} p + 4 \mathcal{K}_{\mu\nu} D\text{ln} p \right. \nonumber \\
&\hspace{60pt} \left.  - 4  D^\perp_\mu \text {ln} p   D^\perp_\nu \text {ln} p \right) \nonumber \\
&+  p^2 (r-r_c)^2  \left( \Omega_{\mu\rho} \Omega_\nu^{\hspace{5pt} \rho} \right).
\end{align}
For reference, the inverse metric of \eqref{second order metric} is presented in Appendix B.

\section{The Entropy Current}

In \cite{Bhattacharyya:2008xc} it was shown that in the fluid-gravity correspondence the entropy current of the dual fluid on the boundary can be mapped into the area current of the black hole event horizon. Thus the second law of thermodynamics is equivalent on a geometrical level to Hawking's area theorem. Here we will follow this general prescription to calculate the entropy current for the Rindler fluid to second order in the gradient expansion. Given the exotic properties of the Rindler fluid, it is clearly of interest to determine whether it behaves consistently with the second law.

First, since metric solution is no longer stationary, the event horizon is dynamical and its location varies in time and space. In order to find $r_h(x^\mu)$  we will need to solve the following equation in the derivative expansion
\begin{align}
g^{A B}  \partial_A (r-r_h(x^\mu)) \partial_B (r-r_h(x^\mu)) = 0 \ .
\end{align}
Using our previous results for the metric, it is straightforward to show at second order
\begin{align}
r_h &= r_c - \frac{1}{p^2} +  \frac{2}{p^3} u^\mu \partial_\mu \text {ln} p - \frac{3}{2p^4} \mathcal{K}_{\alpha \beta} \mathcal{K}^{\alpha \beta} - \frac{1}{2p^4} \Omega_{\alpha \beta} \Omega^{\alpha \beta} \nonumber \\
& - \frac{8}{p^4} D \text {ln} p  D \text {ln} p  + \frac{1}{p^4} D^{\perp \mu} \text {ln} p  D^\perp_ \mu  \text {ln} p + \frac{4}{p^4} D (D \text {ln} p) \ .
\end{align}
We can define a co-dimension 2 hyper-surface by two null normals to the hyper-surface. The ingoing null geodesics $n^A $ and the outgoing null geodesics $\ell^A$ are:
\begin{align}
\ell^\mu = A^\mu, \hspace{10pt} \ell^r = B, \hspace{10pt} n^r = -1, \hspace{10pt} n^\mu = 0 \ .
\end{align}
The unknown functions $ A^\mu$ and $B$ can be found from the following relations:
\begin{align}
\ell^A \ell_A = 0, \hspace{10pt} n^A n_A = 0, \hspace{10pt} \ell_A n^A = -1 \ .
\end{align}
From these conditions we see that there is another freedom in determining $\ell^\mu$ in any order except from the zeroth order. Therefore, we impose the requirement that the vector $\ell^A$ will be the normal vector to foliations of hyper-surfaces that do not intersect with each other. This requirement is called the Frobenius condition and is expressed by the equation:
\begin{align}
v \wedge dv, \hspace{10pt} v_A \equiv g_{A B} \ell^B \ .
\end{align}
This gives us, on the event horizon (putting $r=r_h(x^\mu)$) , the same null normal vector that we get by calculating the normal to the event horizon directly from the equation that defines the normal to the event horizon:  $\ell^A = g^{A B} \partial_B (r - r_h)$. \\
To second order, the vector $\ell^A$ on the horizon is:
\begin{align}
\ell^r &= \ell^\mu \partial_\mu r_h = \frac{2}{p^3} u^\mu \partial_\mu \text {ln} p - \frac{6}{p^4} u^\mu \partial_\mu \text{ln} p u^\rho\partial_\rho \text {ln} p - \frac{2}{p^4} P^{\mu \rho} \partial_\mu \text{ln} p \partial_\rho \text{ln} p + \frac{2}{p^4} u^\mu u^\rho \partial_\mu \partial_\rho \text {ln} p  \\
\ell^\mu &= \frac{1}{p} u^\mu -\frac{1}{2p^3} P^{\mu \rho} P^{\lambda \sigma} \partial_\lambda \partial_\sigma u_\rho - \frac{2}{p^3} \Omega^{\mu \lambda} \partial_\lambda \text {ln} p + \frac{2}{p^3} P^{\mu \nu} \partial_\nu \text {ln} p u^\rho \partial_\rho \text {ln} p - \frac{2}{p^3} P^{\mu \nu}  u^\rho \partial_\nu \partial_\rho \text {ln} p \ .
\end{align}
In order to compute the entropy current we will employ horizon expansion $\theta_{(\ell)}$ along the horizon generator $ \ell^A $
\begin{align}
\theta_{(\ell)} &= \left( g^{A B} + \ell^A n^B + \ell^B n^A \right) \nabla_A \ell_B = \frac{1}{p}\partial_\mu \left(\sqrt{g} p \ell^{(0) \mu} + \sqrt{g} p \ell^{(1) \mu} +\sqrt{g} p \ell^{(2) \mu} \right) \ . \label{expansion}
\end{align}
We can identify the entropy current as the term in the brackets up to an overall factor of  $1/4G$ (which in our units of $16\pi G=1$ is $4\pi$) \cite{Booth:2010kr}.\\
In order to compute the entropy current we therefore need two ingredients: The square root of the metric determinant, and the null generator $\ell^\mu$. We explained how to get the latter. The former can be derived by computing the expansion of the null normal $\tilde{\ell^A}$, where $\tilde{\ell^r} =\ell^r$ and  $\tilde{\ell^u} =\ell^{(0)\mu}$, then we will get only the first term in the brackets of (\ref{expansion}) and we can identify immediately the square root of the determinant of the metric.\\
Combining all the ingredients we get the following result for entropy current:
\begin{align}
S^\mu &= \frac{p \ell^\mu}{4G} \left(1 - \frac{1}{p^2} \left(\mathcal{K}_{\alpha \beta} \mathcal{K}^{\alpha \beta}  - \frac{5}{2} \Omega_{\alpha \beta}  \Omega^{\alpha \beta} + 2 P^{\alpha \beta} \partial_\alpha \partial_\beta \text{ln} p + 2 \mathcal{K} D\text{ln} p - 2 P^{\alpha \beta} \partial_\alpha \text{ln} p \partial_\beta \text{ln} p \right) \right) \ .
\end{align}
Taking the divergence and imposing the Navier-Stokes equations gives
\begin{align}
\partial_\mu S^\mu &= \frac{1}{2Gp} \left(\mathcal{K}_{\alpha \beta} +\frac{1}{p} \left(- 3 \mathcal{K}_{\alpha \beta} u^\mu \partial_\mu \text{ln} p +2\partial_\alpha \text{ln} p \partial_\beta \text{ln} p -2\partial_\alpha \partial_\beta \text{ln} p - 2 \mathcal{K}_\alpha^ \mu \mathcal{K}_{\mu \beta}  -2 \Omega_\alpha ^{\hspace{5pt} \nu} \Omega_{\nu \beta}\right) \right)^2 \ ,
\end{align}
which is clearly non-negative, just as expected from the area increase theorem applied to the Rindler horizon.

\section{Discussion}

In this paper we extended the fluid/gravity correspondence to relativistic Lorentzian geometries.
The fluid is defined on a codimension one timelike hypersurface, embedded in the bulk space-time and equipped with a flat intrinsic metric.
We considered  two types of geometries that are solutions to the vacuum Einstein equations: the Rindler metric and the Taub plane symmetric vacuum. They are found correspond to dual perfect fluids with vanishing and negative energy densities, respectively.
The Rindler geometry is characterized by a causal horizon and one can systematically construct its hydrodynamic derivative expansion.
The Taub geometry, however,  has a timelike naked singularity, indicating pathological behavior.

The dual fluid in the Rindler case has a vanishing energy density as its equation of state. While such an equation of state is rather unusual,
the hydrodynamic expansion up to second order exhibits no pathologies and the gradient of the entropy current
is still positive.
We determined the transport coefficients and verified their consistency with those obtained in the non-relativistic Rindler hydrodynamics.

There are various directions to proceed from here. One may wish to consider the higher curvature corrections the fluid/Rindler relation.
In the AdS/CFT examples this provides information on the holographic relation to field theory deformations as well as at subleading orders in the field theory strong coupling expansion.
Also here, such an analysis is likely to add valuable information on the correspondence. In particular, now the entropy current is no longer the area current \cite{Chapman:2012my} and one needs to define it appropriately and verify the positivity of its gradient.
One may also wish to consider the field theory dual to Rindler geometry and study its properties.
There is at least one candidate non-local dual field theory \cite{Compere:2011dx}, which gives the correct equation of state at the ideal fluid order. One can try to study its
thermal properties and its hydrodynamic limit. This can shed light on holography in asymptotically flat spaces, where the dual field theory is expected to be a non-local
one.

\vskip 1cm
\textbf{Note added}: Sections III and IV contain some overlap with \cite{skenderis} which is posted simultaneously with this paper on the ArXiv.

\appendix
\section{The Connections and Ricci Tensor}
The connections up to 2nd order from the first order metric \eqref {metric1} are:
\begin{align}
\Gamma^{r}{}_{\mu r} &= \frac{1}{2} (3\partial_\mu \text{ln} p + p u_\mu + u_\mu u^\lambda \partial_\lambda \text{ln} p + u^\sigma \partial_\sigma u_\mu ) \nonumber \\
& + (r-r_c) (\partial_\mu u^\lambda \partial_\lambda \text{ln} p + u_\mu P^{\lambda \rho} \partial_\rho \text{ln} p \partial_\lambda \text{ln} p - P^{\lambda \rho} \partial_\rho u_\mu \partial_\lambda \text{ln} p ) \\
\Gamma^{r}{}_{\mu \nu} &= \frac{1}{p} (-\partial_{(\mu} u_{\nu)} + u_{(\mu} \partial_{\nu)} \text{ln} p + \frac{1}{2} p u_\mu u_\nu +  u_\mu u_\nu u^\lambda \partial_\lambda \text{ln} p) \nonumber \\
 &+ p (r-r_c)(3 u_{(\mu} \partial_{\nu)} \text{ln} p + \frac{1}{2} p u_\mu u_\nu +\frac{1}{2} u^\lambda \partial_\lambda (u_\mu u_\nu) + u_\mu u_\nu u^\lambda \partial_\lambda \text{ln} p) \nonumber \\
&+ p^2 (r-r_c)^2 (2 u_{(\mu} \partial_{\nu)} u^\rho \partial_\rho \text{ln} p - P^{\lambda \rho} \partial_\lambda (u_\mu u_\nu) \partial_\rho \text{ln} p ) \nonumber \\
&+ (r-r_c) \left(2\partial_\mu \text {ln} p \partial_\nu \text {ln} p + 2 \partial_\mu \partial_\nu \text {ln} p + 2 u_{(\mu} \partial_{\nu)} u^\rho \partial_\rho \text {ln} p + 2 \partial_{(\mu} u_{\nu)} u^\rho \partial_\rho \text {ln} p \right. \nonumber \\
& \hspace{45pt}  - u_\mu u_\nu u^\lambda \partial_\lambda \text {ln} p u^\rho \partial_\rho \text {ln} p + u^\lambda \partial_\lambda (u_\mu u_\nu) u^\rho \partial_\rho \text {ln} p + u_\mu u_\nu u^\lambda \partial_\lambda u^\rho \partial_\rho \text {ln} p  \nonumber \\
&\hspace{45pt} \left.+ u_\mu u_\nu u^\rho u^\lambda \partial_\rho \partial_\lambda \text {ln} p + 2 u^\lambda \partial_\lambda u_{(\mu} \partial_{\nu)} \text {ln} p+ 2 u_{(\mu} u^\lambda \partial_\lambda\partial_{\nu)} \text {ln} p \right) \\
\Gamma^{\nu}{}_{\mu r} &= \frac{1}{2} p \left( - \partial_\mu u^\nu - u_\mu P^{\nu \lambda} \partial_\lambda \text {ln} p + P^{\nu \sigma} \partial_\sigma u_\mu \right)\\
\Gamma^{\nu}{}_{\mu \sigma} &= \frac{1}{2} \left(p u^\nu u_\mu u_\sigma - 2 u^\nu \partial_{(\mu} u_{\sigma)} + 2 u^\nu u_{(\sigma} \partial_{\mu)} \text {ln} p + 2 u^\nu u_\mu u_\sigma u^\lambda \partial_\lambda \text {ln} p \right) \nonumber \\
& + \frac{1}{2} p^2 (r-r_c) \left(-2 u_{(\sigma} \partial_{\mu)} u^\nu + P^{\nu \lambda} \partial_\lambda(u_\sigma u_\mu) \right) \nonumber \\
&+ p (r-r_c) P^{\nu \rho} \partial_\rho \text {ln} p \left(2 \partial_{(\mu} u_{\sigma)} - 2 u_{(\mu} \partial_{\sigma)} \text {ln} p - 2 u_\mu u_\sigma u^\lambda \partial_\lambda \text {ln} p \right) \nonumber \\
&+ p (r-r_c) P^{\nu \lambda} \left(2 u_{(\sigma} \partial_{\mu)} u_\lambda u^\rho \partial_\rho \text {ln} p - 2 \partial_{(\mu} u_\lambda P^\rho_{\sigma)} \partial_\rho \text {ln} p - 2 \partial_{(\mu} (p u_{\sigma)} P^\rho_\lambda \partial_\rho \text {ln} p) \right. \nonumber \\
&\hspace{65pt} \left.- \partial_\lambda (u_\sigma u_\mu u^\rho \partial_\rho \text {ln} p - 2 u_{(\mu}  P^\rho_{\sigma)} \partial_\rho \text {ln} p)\right)
\end{align}
The Ricci tensor calculated from the 1st order metric \eqref{metric1} up to 2nd order is:
\begin{align}
\label{fullhattedRicci}
\hat{R}_{rr}^{(2)} &= \frac{1}{2} p^2\left(-\partial_\nu u^\mu \partial_\mu u^\nu + P^{\mu \sigma} \partial_\mu u^\nu \partial_\sigma u_\nu \right) \nonumber \\
\hat{R}_{r \mu}^{(2)} &= \frac{1}{2} p (2u_\mu u^\rho \partial_\rho \text{ln} p - 2 P^{\nu \rho} \partial_\nu u_\mu \partial_\rho \text{ln} p + u_\mu P^{\nu \lambda} \partial_\rho \text{ln} p \partial_\nu \text{ln} p - \partial_\nu \partial_\mu u^\nu - u_\mu \partial_\nu u^\nu u^\lambda \partial_\lambda \text{ln} p \nonumber \\
& \hspace{25pt} - u_\mu P^{\nu \lambda} \partial_\nu \partial_\lambda \text{ln} p \nonumber +\partial_\nu u^\nu  u^\sigma \partial_\sigma u_\mu +P^{\nu \sigma} \partial_\nu \partial_\sigma u_\mu ) \nonumber \\
&+ \frac{1}{2} p^3 (r-r_c)(-u_\mu \partial_\sigma u^\nu \partial_\nu u^\sigma + u_\mu P^{\nu \lambda} \partial_\nu u^\sigma \partial_\lambda u_\sigma) \\
\hat{R}_{\mu \nu}^{(2)} &= - \frac{1}{2} \partial_\mu \text{ln} p \partial_\nu \text{ln} p + 2 \partial_\mu \partial_\nu \text{ln} p + 2 u_{(\mu} \partial_{\nu)} u^\rho \partial_\rho \text{ln} p +  \partial_{(\mu} u_{\nu)} u^\rho \partial_\rho \text{ln} p + \frac{3}{2} u^\lambda \partial (u_\mu u_\nu) u^\rho \partial_\rho \text{ln} p \nonumber \\
&+ 2 u_\mu u_\nu u^\lambda \partial_\lambda u^\rho \partial_\rho \text{ln} p + 2 u_\mu u_\nu u^\rho u^\lambda \partial_\lambda \partial_\rho \text{ln} p + 2 u_{(\mu} u^\lambda \partial_\lambda \partial_{\nu)} \text{ln} p - \partial_\sigma u^\sigma \partial_{(\mu} u_{\nu)} - u^\sigma \partial_\sigma \partial_{(\mu} u_{\nu)} \nonumber \\
&+ \partial_\sigma u^\sigma u_{(\mu} \partial_{\nu)} \text{ln} p +u_{(\mu} u^\sigma \partial_\sigma \partial_{\nu)} \text{ln} p + u^\sigma \partial_\sigma u_{(\mu} \partial_{\nu)} \text{ln} p + u_\mu u_\nu \partial_\sigma u^\sigma u^\lambda \partial_\lambda \text{ln} p - \partial_\nu \partial_\mu \text{ln} p \nonumber \\
&- u_{(\mu} \partial_{\nu)} \text{ln} p u^\lambda \partial_\lambda \text{ln} p - \frac{1}{2} u_\mu u_\nu u^\rho \partial_\rho \text{ln} p u^\lambda \partial_\lambda \text{ln} p - \frac{1}{2} u^\rho \partial_\rho u_\mu u^\lambda \partial_\lambda u_\nu - \frac{1}{2} \partial_\mu u^\sigma \partial_\nu u_\sigma  \nonumber \\
& + \frac{1}{2} P^{\sigma \rho} \partial_\sigma u_\nu \partial_\rho u_\mu - \frac{1}{2} P^{\sigma \rho} \partial_\rho (u_\mu u_\nu) \partial_\sigma \text{ln} p + \frac{1}{2} u_\mu u_\nu P^{\sigma \lambda} \partial_\sigma \text{ln} p \partial_\lambda \text{ln} p \nonumber \\
&+ p^2 (r-r_c) \left( -\partial_\sigma u_{(\nu} \partial_{\mu)} u^\sigma - u_{(\nu} \partial_{\mu)} \partial_\sigma u^\sigma + \frac{1}{2} \partial_\sigma u^\sigma u^\lambda \partial_\lambda (u_\mu u_\nu) + u_{(\mu} \partial_{\nu)} u^\sigma \partial_\sigma \text{ln} p \right. \nonumber \\
&\hspace{65pt}  - \frac{1}{2} P^{\sigma \rho} \partial_\rho (u_\mu u_\nu) \partial_\sigma \text{ln} p + \left. u_\mu u_\nu u^\lambda \partial_\lambda u^\sigma \partial_\sigma \text{ln} p + \frac{1}{2} \partial_\mu u^\lambda \partial_\nu u_\lambda - P^{\sigma \rho} \partial_\rho u_\nu \partial_\sigma u_\mu \right) \nonumber \\
&+ \frac{1}{2} p^4 (r-r_c)^2 \left(- u_\mu u_\nu \partial_\lambda u^\sigma \partial_\sigma u^\lambda + u_\mu u_\nu P^{\sigma \rho} \partial_\sigma u^\lambda \partial_\rho u_\lambda\right).
\end{align}
\section {Inverse metric}
The inverse metric of (\ref{second order metric}) to the 2nd order is:
\begin{align}
g^{r r} &= \frac{1}{p^2} \left(1+p^2(r-r_c)\right) -\frac{2}{p}(r-r_c)u^\lambda \partial_\lambda \text{ln} p \nonumber \\
&-\frac{1}{2} (r-r_c)^2 \left( \mathcal{K}_{\alpha \beta} \mathcal{K}^{\alpha \beta} - 6 P^{\rho \lambda}  \partial_\lambda \text{ln} p  \partial_\rho \text{ln} p\right) - \frac{1}{2} p^2 (r-r_c)^3 \Omega_{\alpha \beta} \Omega^{\alpha \beta} \nonumber \\
&- \frac{2}{p^2} (r-r_c) \mathcal{K}_{\alpha \beta} \mathcal{K}^{\alpha \beta} \\
g^{r \mu} &= \frac{1}{p} u^\mu - 2(r-r_c) P^{\mu \lambda} \partial_\lambda \text{ln} p + \frac{1}{2} p (r-r_c)^2 \left(4 \Omega^{\mu \sigma}  \partial_\sigma \text{ln} p + P^{\mu \rho} P^{\lambda \sigma} \partial_\lambda  \partial_\sigma u_\rho \right) \nonumber \\
& - \frac{1}{p} (r-r_c) \left(- P^{\mu \rho} P^{\lambda \sigma} \partial_\lambda  \partial_\sigma u_\rho + 2\mathcal{K}^{\mu \sigma}\partial_\sigma \text{ln} p  - 2\Omega^{\mu \sigma} \partial_\sigma \text{ln} p\right) \\
g^{\mu \nu} &= P^{\mu \nu} - \left(2\mathcal{K}^\mu_\rho \mathcal{K}^{\nu \rho}  + 4 \mathcal{K}^{\rho (\mu} \Omega_{|\rho|}^{\hspace{5pt} \nu)}  + 4  \Omega^\mu_{\hspace{5pt} \rho} \Omega^{\rho \nu}
 + 4  P^{\mu \alpha} P^{\nu \beta} \partial_\alpha  \partial_\beta \text{ln} p + 4 \mathcal{K}^{\mu\nu} D\text{ln} p \right. \nonumber \\
&\hspace{55pt} \left.  - 4  D^{\perp \mu} \text {ln} p   D^{\perp \nu} \text {ln} p \right)(r-r_c) -p^2 (r - r_c)^2 \Omega^\mu_{\hspace{5pt} \rho} \Omega^{\rho \nu}
\end{align}

\section*{Acknowledgements}

The work is supported in part by the Israeli Science Foundation center
of excellence.

\end{document}